\documentclass[]{aastex701}

\usepackage{fontawesome}

\begin{document}

\title{\texttt{Alfred}: A Flexible, User-Friendly Python Package for Exoplanet Confirmation}

\author[orcid=0009-0002-2757-4138]{Maxwell A. Kroft}
\email[show]{mkroft@wisc.edu} 
\affiliation{Department of Astronomy, University of Wisconsin--Madison, 475 N. Charter Street, Madison, WI, 53706, USA}

\begin{abstract}

In the era of TESS, exoplanet follow-up and confirmation have become a massive community effort. With only about an eighth of the over 8,000 TESS exoplanet candidates confirmed, astronomers need flexible software to allow them to characterize many different kinds of systems with data from multiple telescopes and instruments. In this work, I present \texttt{Alfred} (the Awesome Library For Robust Exoplanet Detection), an open-source Python package for exoplanet detection and confirmation. \texttt{Alfred} is designed to be flexible and easy to use, allowing users to fit transit and radial velocity data for an arbitrary system of exoplanets. The code is released under an MIT license on \href{https://github.com/maxkroft/Alfred}{GitHub \faGithub} and is installable via pip, with \href{https://alfred-exoplanets.readthedocs.io/en/latest/index.html#}{documentation} under active development.

\end{abstract}




\section{Description of \texttt{A\lowercase{lfred}}}

\texttt{Alfred} wraps together all of the initialization, modeling, fitting, and output inherent in exoplanet transit and radial velocity (RV) fitting, streamlining the process for the user. It is built on the \texttt{emcee} \citep{emcee} Markov chain Monte Carlo (MCMC) sampler, the \texttt{batman} \citep{batman} planetary transit model, and the \texttt{isochrones} \citep{isochrones} stellar model interpolator, as well as a numerical solver for Kepler's equation to model RVs. It also comes with precomputed quadratic limb darkening grids generated using \texttt{ExoCTK} \citep{exoctk}.

Key features of \texttt{Alfred} v1.1, which is used for this publication, include:

\begin{itemize}
    \item Generalized transit and RV fitting for any number of planets in a system. Individual planets are not required to be detectable in every data set.
    \item Simultaneous light curve detrending using the \texttt{celerite2} \citep{celerite1,celerite2} damped harmonic oscillator Gaussian process kernel.
    \item Optional simultaneous stellar parameter and SED fitting.
    \item Simple transit timing variation fitting, treating each transit center time as a free parameter.
    \item User-defined priors for any fit parameter, as well as the option to fix any parameter to a specific value.
    \item Human-readable initialization files, each with a tailored GUI to help users set them up correctly.
    \item Support for photometry from multiple instruments (with different limb darkening) or with different exposure times, as well as RVs from multiple instruments.
    \item Automatic plot generation of MCMC chain distributions and best-fit models to the data.
    \item The ability to load previous runs to remake plots or use raw chains.
\end{itemize}

\begin{figure}[t]
    \centering
    \includegraphics[width=\linewidth]{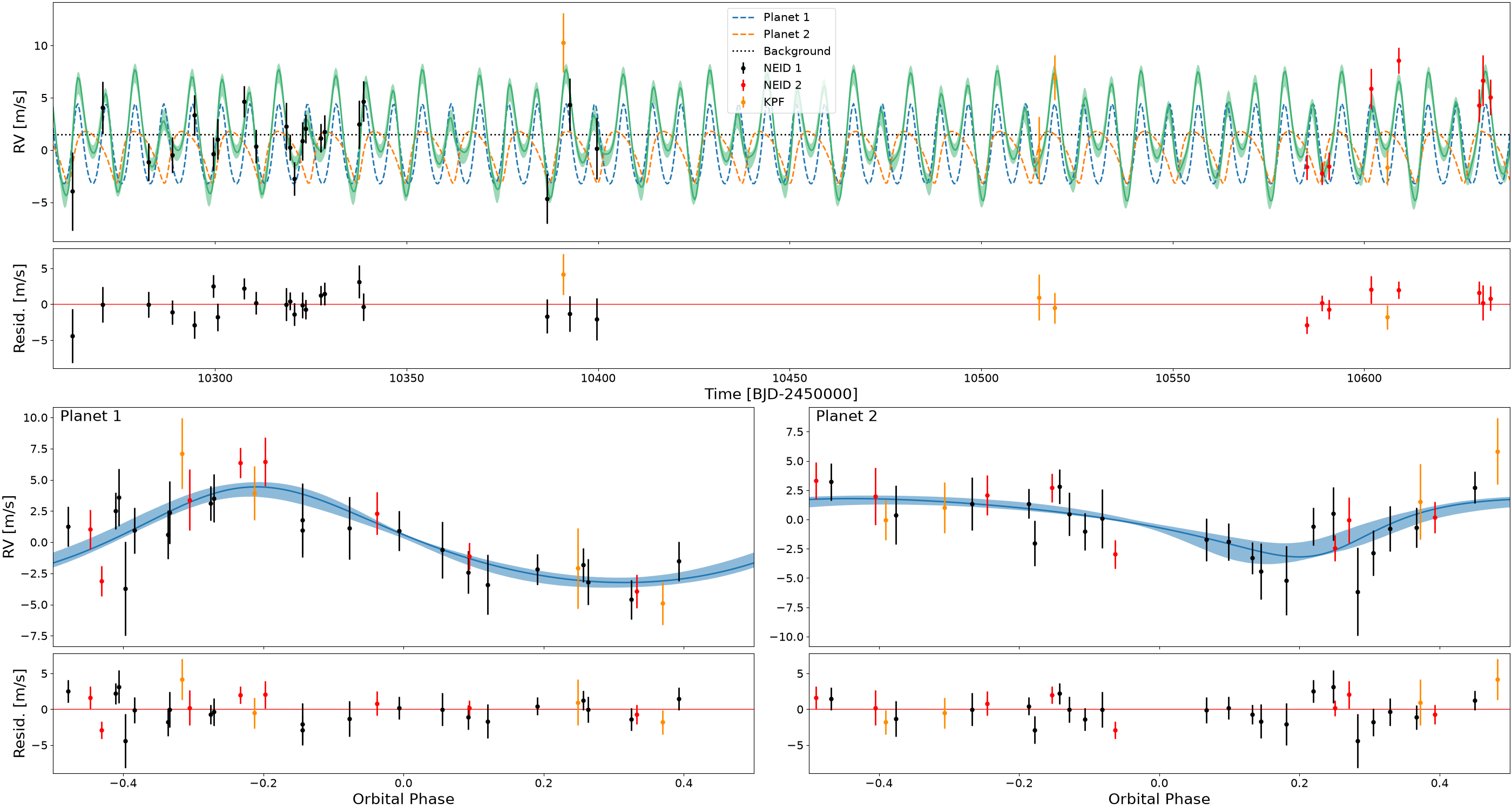}
    \caption{An example RV fit with \texttt{Alfred} on the TOI-6054 system \citep{Kroft_2025}, with Gaussian priors on the planets' periods and conjunction times. This fit demonstrates \texttt{Alfred}'s ability to fit multiple planets and multiple datasets simultaneously, as well as handle eccentricity and user-defined priors. This plot is the default output from an RV fit with \texttt{Alfred.}}
    \label{fig:rv}
\end{figure}

\section{Usage Examples}

\texttt{Alfred} was designed with Python notebooks in mind, but can easily be run with standard scripts as well. Data, initialization files, and \texttt{Alfred} outputs are all stored in a directory specific to the exoplanet system, which the user can create with \texttt{Alfred} using built-in functions. Data and initialization parameters are read in via an \texttt{ExoSystem} object, which has class functions to run fits, generate plots, or perform other analyses.

\texttt{Alfred} has been used two confirm exoplanets in two scientific papers so far: TOI-6054b and c \citep{Kroft_2025}, which used a private, in-development version of the software, and GJ 523b \citep{Kroft_2026}. Figure \ref{fig:rv} shows an example RV fit for the TOI-6054 system performed with the current version of \texttt{Alfred}. This is a system with two eccentric sub-Neptunes, and I use the NEID data presented in the original publication (split into two RV eras defined by the NEID science team) as well as KPF data that has since become public. As this is an RV-only fit, Gaussian priors were placed on the period and time of conjunction for both planets, based off of the results of transit fitting. The plot shown is the default output from \texttt{Alfred}.

\texttt{Alfred} is available on \href{https://github.com/maxkroft/Alfred}{GitHub \faGithub} and is installable via pip under the name \texttt{alfred-exoplanets}. Documentation, including installation instructions, are available \href{https://alfred-exoplanets.readthedocs.io/en/latest/index.html#}{here}. I am still actively developing the software and documentation, and plan to add many more features in the future. Questions, comments, and suggestions for the code are welcome and appreciated on the GitHub. Finally, I would like to thank Tayt Armitage for designing the \texttt{Alfred} logo.

\software{\texttt{astropy} \citep{astropy:2013,astropy:2018,astropy:2022}, \texttt{numpy} \citep{numpy}, \texttt{matplotlib} \citep{matplotlib}, \texttt{corner} \citep{corner}, \texttt{scipy} \citep{scipy}, \texttt{customtkinter} \citep{customtkinter}, \texttt{astroquery} \citep{astroquery}}



\bibliography{bib}{}
\bibliographystyle{aasjournalv7}



\end{document}